\documentclass{article}

\usepackage{arxiv}
\usepackage{amsmath}
\usepackage[utf8]{inputenc} 
\usepackage[T1]{fontenc}    
\usepackage{hyperref}       
\usepackage{url}            
\usepackage{booktabs}       
\usepackage{amsfonts}       
\usepackage{nicefrac}       
\usepackage{microtype}      
\usepackage{cleveref}       
\usepackage{lipsum}         
\usepackage{graphicx}
\usepackage{natbib}
\usepackage{doi}
\usepackage{multirow}
\usepackage[table]{xcolor}
\usepackage{algorithmic}
\usepackage{algorithm}
\newtheorem{theorem}{Theorem}
\newtheorem{lemma}{\textbf{Lemma} }
\newtheorem{proof}{\textbf{Proof} }
\newcommand{\argmin}[2]{\operatornamewithlimits{argmin}\limits_{#1}{\;\;#2}}

\newcommand{\mino}[2]{\operatornamewithlimits{min}\limits _{#1}{\;\;#2}}
\def \*#1{\mathbf{#1}}   
\def \r{\mathbf{r}}

\title{PSTAIC regularization for 2D spatiotemporal  image  reconstruction}

\newif\ifuniqueAffiliation
\uniqueAffiliationtrue

\ifuniqueAffiliation 
\author{ Deepak G Skariah \\
	Department of Electrical Engineering \\
	IISc \\
	Bengaluru, 560012 \\
	\texttt{deepaks@iisc.ac.in} \\
	\And
	Muthuvel Arigovindan \\
	Department of Electrical Engineering\\
	IISc\\
	Bengaluru,560012\\
	\texttt{mvel@iisc.ac.in} \\
}
\else
\usepackage{authblk}

\newbox{\orcid}\sbox{\orcid}{\includegraphics[scale=0.06]{orcid.pdf}} 
\author[1]{%
	\href{https://orcid.org/0000-0000-0000-0000}{\usebox{\orcid}\hspace{1mm}David S.~Hippocampus\thanks{\texttt{hippo@cs.cranberry-lemon.edu}}}%
}
\author[1,2]{%
	\href{https://orcid.org/0000-0000-0000-0000}{\usebox{\orcid}\hspace{1mm}Elias D.~Striatum\thanks{\texttt{stariate@ee.mount-sheikh.edu}}}%
}
\affil[1]{Department of Electrical Engineering, IISc Bangalore}
\affil[2]{Department of Electrical Engineering, IISc Bangalore}
\fi

\renewcommand{\shorttitle}{\textit{PSTAIC} Algorithm}

\hypersetup{
pdftitle={A template for the arxiv style},
pdfsubject={q-bio.NC, q-bio.QM},
pdfauthor={David S.~Hippocampus, Elias D.~Striatum},
pdfkeywords={First keyword, Second keyword, More},
}

\begin{document}
\maketitle

\begin{abstract}
 We propose a model for restoration of spatio-temporal TIRF images based on infimal decomposition regularization model named STAIC proposed earlier. We propose to strengthen the STAIC algorithm by enabling it to estimate the relative weights in the regularization term by incorporating  it as part of the optimization problem. We also design an iterative scheme which alternatively minimizes the weight and image sub-problems. We demonstrate the restoration quality of this regularization scheme against other restoration models enabled  by similar weight estimation schemes.
\end{abstract}


 \section{Introduction}
 
 Image restoration is an inverse problem where a better quality image is estimated from a corrupted measurement by assuming knowledge of  the image  formation forward model.  We consider the problem of restoration of  spatio-temporal TIRF microscopy images. One of the popular approaches to image restoration is to model the restored image $g$ as   solution of an optimization model. This approach is known as regularized image restoration and it  models the optimization cost as sum of a data fitting cost and a regularization cost. Formally, if $m$ is the measured image, the regularized image restoration problem takes the following form:
 
 \begin{equation} \label{Equation  : Regularization}
	{{g}}_{opt}  	 = \argmin{{g}}{G({ g,m} )  + \lambda  R({g})}
\end{equation} 
 
 where $G({ g,m}) $ is the data fitting cost and   $R(g)$ is the regularization cost. The regularization  parameter $\lambda$   controls the relative importance of  data fitting term as against the regularization term in the overall cost. 
 The data fitting cost  which ensures that the image estimate is not too far away from the measurement made by the imaging device. The regularization cost ensures that the estimate $g_{opt}$ has the common features of the class of image being estimated. In other words, $R(g)$ captures the image statistics of an ideal image estimate. 
 In   our  paper  \cite{skariah2024staic}, we discussed the  problem of designing an algorithm for recovering a better quality estimate from a blurred noisy  spatio-temporal image from a TIRF microscope.  In this paper, we will take a closer look at the same problem but from  a   parameter selection point of view.  Recalling the STAIC optimization problem  from \cite{skariah2024staic}:
 \begin{align}  
 	\label{Equation:STAICMODELBASIC} 
 	&	(g_{opt},v_{opt}) =  \argmin{g,v}{G(h,g,m) +\bar{S}(g,v,\alpha_s,\alpha_t)}\\  &\text{ where } 
 	\bar{S}(g,v,\alpha_s,\alpha_t) =  \alpha_s R_1(g,v) +\alpha_t R_2(v) \nonumber
 \end{align}  
  where $R_1(g,v) = \sum_{i=1}^{n_F}\|A*(g_i-v_i)  \|_{1,2}$ and $	R_2(v) = \|(M*v) \|_{1,F}$. Here $M$ and $A$ are two linear operators defined as shown below:
  \begin{equation*}
  	A(\*k)=\begin{bmatrix}
  		\bar{d}_{xx}(\*k)   \\    
  		\bar{d}_{xy}(\*k)   \\
  		\bar{d}_{yx}(\*k)      \\
  		\bar{d}_{yy}(\*k)   \\
  		\delta(\*k)
  	\end{bmatrix}, 	M(\*r)=\begin{bmatrix}
  		d_{xx}(\*r)       & d_{xy}(\*r)  & d_{xt}(\*r) \\
  		d_{yx}(\*r)       & d_{yy}(\*r)  & d_{yt}(\*r) \\
  		d_{tx}(\*r)       & d_{ty}(\*r)  & d_{tt}(\*r)  
  	\end{bmatrix}
  \end{equation*}
  
  Here, $\bar{d}_{xx}(\*k),\bar{d}_{xy}(\*k),\bar{d}_{yy}(\*k)$ are discrete filters implementing 2D second order derivatives and $\delta(\*k)$ is the 2D Kronecker delta. In addition  $d_{xx}(\*r),d_{xy}(\*r), d_{xt}(\*r)$,$d_{yy}(\*r),d_{yt}(\*r),d_{tt}(\*r)$ are   discrete filters implementing the 3D second order derivatives. Also  $M(\*r)$ is the 3D Hessian operator  encapsulating   all the second order derivative operators where the directions are $(x,y,t)$. It may also be noted that $\*k$ and $\*r$ represents 2D and 3D pixel indices respectively. 
  The user defined parameters $\alpha_s$, $\alpha_t$ helps control the relative strength of $R_1( \cdot, \cdot )$ and $R_2( \cdot)$ in the regularization function $S(g,\alpha_s,\alpha_t)$.  
  Here $g$  and $v$ are 2DT images, and $g_i$  is the $i^{th}$ 2D time frame of    of the 2DT image $g$.  The data fitting term employed in STAIC regularization  is 
  \begin{equation}
  	\label{Equation : datafitting }
  	G(h,g,{m})=\frac{1}{2}\sum_{i=1}^{n_F}\|(h*g_i)-m_i\|_{F}^2
  \end{equation}

  It also may be noted that the blurring happens frame wise as we are considering a 2D signal observed over time. 
 In the STAIC model, the parameters $\alpha_s$ and $\alpha_t$ where selected by the algorithm  user and was selected before running the iterative ADMM scheme. The parameters determines  the relative importance of the two terms $R_1$ and $R_2$ in  overall regularization term $\bar{S}(g,v,\alpha_s,\alpha_t)$. In this paper, we address the problem of finding the signal estimate along with weighing parameters of the STAIC regularization ($\alpha_s,
 \alpha_t$)  as a single joint estimation problem.    The choice of these weights determine the emphasis laid on spatial vs spatio-temporal regularization on the two  components of the restored signal created as a result of underlying infimal decomposition  components affecting the final restoration quality.  The selection of relative weights in the regularization term is a difficult problem due to the lack of knowledge of motions locally. Our proposed formulation poses this as a joint optimization problem where $\alpha_s$  and $g$ are treated as variables to be estimated. 
 
 Our  contributions  in this paper can be summarized as follows:
 \begin{itemize}
 	\item We begin by introducing the optimization model where the weights of the two regularization terms are modeled as optimization variables.
 	
 	\item We describe the alternating minimization scheme  used in minimization of the resultant optimization problem.
 	
 	\item We also study the properties of the resultant sub-problems  of the optimization problem  and analyzes its convexity and differentiability properties.
 	\item Finally we demonstrate the effectiveness of the resultant parameter selection scheme using experiments on the TIRF spatio-temporal signal restoration problem. 
 \end{itemize}

 \subsection{Organization of the Paper}
 
 We start by presenting the notations and mathematical preliminaries needed to understand the algorithm in \Cref{Section : Notation}. We motivate the  joint estimation model  of finding the weights and the restored image estimate in \Cref{Chapter_4:Motivation}. This is followed by description of the actual non-convex optimization  model that achieves this joint estimation in \Cref{Chapter_4:PSTAIC}.  This section also discusses how  the weights sub problem that estimates the weight $\alpha_s$ is optimized. A discussion of the image estimation problem and the algorithm is followed next in \Cref{Chapter_4:PSTAIC_ALGORITHM}. In \Cref {Section : ADMM subproblems} we discuss the details of sub-problems that arise out of our algorithm. Finally the experimental validation of the model  for the TIRF signal estimation problem is demonstrated for a simulated data-set in \Cref{Chapter_4:EXPERIMENTS}. 
 
 \section{Notations and mathematical preliminaries}{\label{Section : Notation}}
 
 \begin{enumerate}
 	
 	\item Images are represented by lower case English alphabets. For example $g$.
 	
 	\item
 	In the discussion we will use the idea of vector valued images  often refereed to as vector images.  Vector images are discrete 2D arrays where each pixel
 	location has a vector quantity. It is denoted by lower-case bold-faced letter
 	with a bold-faced lower-case letter as an argument.  For example,
 	${\mathbf v}({\mathbf r})$  is a vector image with ${\mathbf r}=[x\;y]^\top $ representing a 2D pixel
 	location. Depending on the context, the symbol denoting the pixel location
 	may be omitted.
 	
 	\item
 	For a vector image, 	${\mathbf v}({\mathbf r})$,  $\|{\mathbf v}\|_{1,2}$
 	denotes  $\|{\mathbf v}\|_{1,2}=\sum_{\mathbf r}\|{\mathbf v}({\mathbf r})\|_{2}$. It is
 	the  sum of  pixel-wise $l_2$ norms, where $\sum_{\mathbf r}$ denotes the sum
 	across pixel indices. The bound of sum is the first to last pixel location in this notation.
 	The norm $\|{\mathbf v}\|_{1,2}$ is a composition of  norms and is often refereed to as a mixed norm.
 	
 	\item In a scalar image having multiple frames, we use the subscript notation to refer to a particular frame number. Example $g_i$ refers to frame number $i$ of the spatio-temporal image $g$.
 	\item Index $\mathbf{r}$ is used to refer to a spatio-temporal image (2DT). Index $\mathbf{k}$ is used to refer to a 2D image. Let $\delta(\*k)$
 	and $\delta(\*r)$ represent 2D and 3D Kronecker delta respectively.
 	\item $\|\mathbf{g}\|_{1,\kappa} =  \sum_{\mathbf{r}} \|\mathbf{g}(\mathbf{r}) \|_{\kappa}$ 
 	with the definition of 
 	$\small \|\mathbf{y}\|_{\kappa} =   \sqrt{\kappa^2 (y_1^2 +y_2^2) +y_3^2}$.  
 	Further we have that 
 	$\|\mathbf{g}\|_{1,1/\kappa} =  \sum_{\mathbf{r}} \|\mathbf{g}(\mathbf{r}) \|_{1/\kappa}$ 
 	and
 	$\|\mathbf{x}\|_{1/\kappa} =   \sqrt{ y_1^2 +y_2^2+ \kappa^2 y_3^2}$.  Here $\kappa$ is a parameter.
 
 \end{enumerate}

 \section{Motivation} \label{Chapter_4:Motivation} 
 
 In image processing, one can find many instances of estimation problems where there is an auxiliary unknown variable in addition to the central variable being estimated. A classical example is the popular k-SVD \cite{aharon2006k} algorithm where the central problem is to find a sparse representation of a  given signal with respect to a   dictionary. Here you are expected to estimate the sparse representation along with the dictionary against which the representation is defined with the help of a given large dataset representing the signal class.  The k-SVD algorithm  approaches this problem as a joint estimation problem where the dictionary and the sparse representation is estimated simultaneously. Another example of this strategy  is observable in blind image deblurring problems where the blur and the deblurred signal are estimated from the observed blurred signal by a joint estimation \cite{levin2006blind} problem.  A more recent example of joint estimation   in imaging inverse problems   is the {\em COROSA} \cite{corosa} algorithm where a sum of norms regularization was designed for restoration of images where the weighting of  two terms in the sum formulation was estimated along with the image.   In {\em COROSA} algorithm, the optimization model solved took the following form:
 \begin{equation}
 	\argmin{\*f,0 \leq \pmb{\alpha}(\*r) \leq 1\;\; \forall \*r}{D(\*f) + \sum_{\*r}\bigg(\pmb{\alpha}(\*r).  R_1(\*f(\*r)) +  \big(1-\pmb{\alpha}(\*r)\big) .R_2(\*f(\*r))\bigg) + \sum_{\*r}P(\pmb{\alpha}(\*r)) }
 \end{equation}
 
 \begin{enumerate}
 	\item Here $D(\*f)$ is the data fitting term
 	\item $ \displaystyle \sum_{\*r}\bigg(\pmb{\alpha}(\*r).  R_1(\*f(\*r)) +  \big(1-\pmb{\alpha}(\*r)\big) .R_2(\*f(\*r))\bigg)$ is the regularization term designed as a sum of 2 different regularization functions $R_1(\cdot)$ and $R_2(\cdot)$. It is a weighted sum of the form $\alpha_1 R_1(\cdot) + \alpha_2 R_2(\cdot)$ . (Note : This is an instance of a sum of norms regularization \cite{lindsten2011clustering} where both the terms of regularization act on the same variable $\*f$ and do not involve an infimal decomposition as observed in STAIC.)
 	\item $\displaystyle \sum_{\*r}P(\pmb{\alpha}(\*r))=-\sum_{\*r}\log(\pmb{\alpha}(r)(1-\pmb{\alpha}(\*r)))$ is a penalty term that ensures that weights $\pmb{\alpha}(r)$   do not subject to  rapid switching \cite{corosa} from a value close to 0 to a value close to 1 (or vice versa) from one iteration to the next.
 	
 \end{enumerate}
 Here $\*r$ is the pixel index and $\sum_{\*r}$  represents  traversal of  all the pixels of the image in the summation. This optimization model was solved by  using an block coordinate descent (BCD) scheme. Note that, here $\pmb{\alpha}$ the weight variable to be estimated is a vector variable having the same size as $\*f$. If $\*f \in \mathbb{R}^n$ then $\pmb{\alpha} \in \mathbb{R}^n$ as well.  This model demonstrates the potential to reduce user overhead in weighting parameter selection in regularization  by resorting to tools in optimization.
 As opposed to treating relative weights as hyper-parameters ( borrowing a machine learning terminology to refer to tuned model parameters), its possible to incorporate the weights as part of the model itself.
 At the same time, there is need to address the fundamental difference in infimal decomposition based  regularization (which by itself is posed as an optimization based regularization function) against a simple sum of functions model of regularization design which is more simple to deal with and interpret. How we address this challenge of infimal decomposition in the signal variable $\*f$ while incorporating the weights $(\alpha_s,\alpha_t)$ is explained in the next section.

 \section{ PSTAIC: A composite optimization model for weights  and image }\label{Chapter_4:PSTAIC}

 In this section, we will introduce a model for joint optimization of spatio-temporal image $g$ and relative weights $(\alpha_s,\alpha_t)$ which we name PSTAIC (Parametric STAIC) indicating that it also estimates parameters/weights for the STAIC regularization function.   We take a re-look at our  STAIC optimization model  from  \cite{skariah2024staic}  as we attempt to design an algorithm serving the the target of simultaneously estimating the two components. Recalling the  STAIC regularization model:
 \begin{align*}\small
 	J(\*f,\lambda, \alpha_s,\alpha_t) = & \frac{1}{2}\sum_{i=1}^{n_F}\|(\*h*\mathbf{f}_i) -m_i \|_{1,F}^2+ \lambda \big(
 	\sqrt{2} \alpha_s\sum_{i=1}^{n_F}\sum_{\mathbf{k}}\|A_s \big((\mathbf{T_s}*\*f_i) \big)(\*k)\|_{2}\\&+\alpha_t  \|\mathbf{T_t}*\*f\|_{1,2}\big) +	\mathcal{B_C}
 	\*f) 
 \end{align*} 
 where 
 \begin{align*}
 	{\mathbf h} & = [h(\mathbf{k})\;\;0 ],  \\
 	{\mathbf T}_{\mathbf s} & = \left[\begin{array}{cc}
 		d_{xx}(\mathbf{k}) & 0   \\ 
 		d_{xy}(\mathbf{k}) & 0  \\
 		d_{yx}(\mathbf{k}) & 0  \\
 		d_{yy}(\mathbf{k}) & 0  \\
 		\delta(\mathbf{k})    & 0 \\
 		0         &  d_{xx}(\mathbf{k}) \\
 		0         &  d_{xy}(\mathbf{k}) \\	
 		0         &  d_{yx}(\mathbf{k}) \\	
 		0         &  d_{yy}(\mathbf{k}) \\
 		0         &   \delta(\mathbf{k})
 	\end{array}   \right],  \;\; 
 	{\mathbf T}_{\mathbf{t}}=\left[ \begin{array}{cc}
 		0 & d_{xx}(\*r)   \\ 
 		0 & d_{yy}(\*r)   \\
 		0 & d_{xy}(\*r)   \\ 
 		0 & d_{yx}(\*r)   \\ 
 		0 & d_{xt}(\*r)   \\
 		0 & d_{tx}(\*r)   \\
 		0 & d_{yt}(\*r)   \\
 		0 & d_{ty}(\*r)   \\
 		0 & d_{tt}(\*r)  
 	\end{array} \right] 
 \end{align*}
 
 \begin{align} \label{Definition : Matrix A_s}
 	&{A}_s =  \small \left[ \begin{array}{cccccccccc}
 		1/\sqrt{2} & 0 & 0 & 0 &0 & -1/\sqrt{2} & 0 & 0 & 0  & 0 \\
 		0&1/\sqrt{2} & 0 & 0 & 0 &0 & -1/\sqrt{2} & 0 & 0& 0    \\
 		0&0&1/\sqrt{2} & 0 & 0 & 0 &0 & -1/\sqrt{2} & 0& 0   \\ 
 		0&0&0&1/\sqrt{2} & 0 & 0 & 0 &0 & -1/\sqrt{2} & 0 \\
 		0&0&0&0&1/\sqrt{2} & 0 & 0 & 0 &0 & -1/\sqrt{2} 
 	\end{array}  \right]
 \end{align}
 \text{ and }
 
 \begin{equation*}
 	\mathcal{B_C}(\mathbf{f}) =
 	\begin{cases}
 		\text{0} &\quad\text{if } \*f \in \mathcal{C} \text{ (The set }\mathcal{C} \text{ is the bound on pixel values) } \\
 		\infty &\quad\text{otherwise} 
 	\end{cases}
 \end{equation*}  
 Here $\lambda$  is the regularization parameter and the  two weighting parameters $\alpha_s$ and $\alpha_t$ were also user tuned parameters selected before running the optimization algorithm. We restricted $\alpha_t=1-\alpha_s$ during experiments to ensure the ease of tuning. The choice of $\alpha_s$ determines the strength of the pure spatial regularization term against the combined spatio-temporal regularization. Hence, value of $\alpha_s$  is central to the restoration quality of the estimated spatio-temporal signal. 
 
 In our PSTAIC model, we intend to strengthen our existing restoration scheme (STAIC)  by enabling it to estimate the signal $\*f$ along with the relative weight term $\alpha_s$. One important  observation here  is the fact that our STAIC regularization itself is equipped to handle spatio-temporal variability (presence of   motion) in the signal due to its design using infimal convolution as opposed to   sum of norms regularization approach which is spatially uniform in action.  In {\em COROSA} described in motivation section (which employs sum of norms), spatial sensitivity (ability to change behavior spatially)   is achieved   by employing  spatially varying weights as part of estimation model.
 We propose to design an estimation model with weight $\alpha_s$ as a model variable but at the same time recognizing that STAIC has inbuilt spatial sensitivity built into it  courtesy the infimal convolution approach to its design. 
 
 \noindent   In summary, the design goals of PSTAIC model can be summarized as follows:
 
 \begin{enumerate}
 	\item Model should be  sensitive to  spatial  variability  inherent to infimal convolution approach to regularization design.
 	\item Model should have the  weighing parameter $\alpha_s$ as part of estimation model.
 \end{enumerate}
 How we achieve the design goals through optimization model redesign  is described next.

 \subsection{PSTAIC Optimization Model}
 PSTAIC model is designed from  the STAIC model   by assigning   $\alpha_s$ as a model variable along with the image variable $\*f$. The resultant  PSTAIC optimization problem can be stated as follows:
 \begin{align}
 	\mino{\*f,0 \leq \alpha_s \leq 1}{ \frac{1}{2}\sum_{i=1}^{n_F}\|(\*h*\mathbf{f}_i) -m_i \|_{1,F}^2+ 
 		\lambda \big(\sqrt{2} {\alpha_s } \sum_{i=1}^{n_F}\sum_{\*k}\|A_s\big((\mathbf{T_s}*\*f_i)(\*k) \big)\|_{F}  +   (1- {\alpha_s })\|(\mathbf{T_t}*\*f)  \|_{2}\big)  +	\mathcal{B_C}(\*f)+P_{\tau}( {\alpha_s})}
 \end{align}
 
 Observe that we have restricted $\alpha_t$ to take value $1-\alpha_s$ while constraining the relative weight $\alpha_s$ to take a value between 0 and 1 . 
 The PSTAIC cost function in two variables $(\*f, \alpha_s)$ is observable from the above optimization problem which we refer to as $H(\*f,{\alpha_s} )$ in subsequent discussion.   
 \begin{align*}\small
 	H(\*f, {\alpha_s} ) = &\frac{1}{2}\sum_{i=1}^{n_F}\|(\*h*\mathbf{f}_i) -m_i \|_{1,F}^2+ 
 	\lambda \big(\sqrt{2} {\alpha_s } \sum_{i=1}^{n_F}\sum_{\*k}\|A_s\big((\mathbf{T_s}*\*f_i)(\*k) \big)\|_{F}  +   (1- {\alpha_s })\|(\mathbf{T_t}*\*f)  \|_{1,F}\big)  +	\mathcal{B_C}(\*f)+P_{\tau}( {\alpha_s})
 \end{align*}  
 
 It may be observed that the cost $	H(\*f, {\alpha_s} )$ is almost similar to the STAIC cost formulation with two fundamental differences. The first major difference is that the weight $\alpha_s$ is absorbed into the optimization cost as a model variable that needs to be optimized. The second major difference is that we have introduced a function $P_{\tau}( {\alpha_s})$ involving the weight variable $\alpha_s$ to ensure that the estimated weights display a certain behavior as described in the next subsection. The definition of $P_{\tau}( {\alpha_s})$   is similar to the term in sum of norms formulation discussed in motivation. Here the  function acts on a 1-D variable $\alpha_s$ where as we have an  n-dimensional variable $\pmb{\alpha}$ in motivation.

 \subsection{Understanding the term $P_{\tau}( {\alpha_s})$}  \label{Subsection: P Term describe}
 The weight regularization term $P_{\tau}( {\alpha_s})= -\tau\log\big(\alpha_s(1-\alpha_s)\big)$  serves multiple purposes in the PSTAIC optimization formulation. To help interpret its role, we first appeal to the concept of a 'barrier' function  \cite{luenberger2016penalty} from nonlinear optimization theory.  Consider the following constrained optimization problem: 
 \begin{align*}
 	\min_x\;\; f_1(x)\\ \text{subject to } g_1(x) \leq 0 \\g_2(x) \leq 0
 \end{align*} 
 One  approaches in numerical optimization to solve such a constrained optimization problem is to solve a sequence of unconstrained  optimization problems of the following form:
 \begin{align}
 	\mino{x}{f_1(x)-\big(log(- g_1(x))+log(- g_2(x))\big)}
 \end{align} 
 Here the term $-\log(-g_i(x) )$ is referred to as a 'barrier' term. This is because of the following fact: 
 \begin{equation}
 	-\log(-g_i(x)) =
 	\begin{cases}
 		\text{finite } &\quad\text{if } g_i(x) \leq 0 \\
 		\infty &\quad\text{otherwise} 
 	\end{cases}
 \end{equation}  
 This means that $-\log(-g_i(x))$ acts as a barrier to $x$ taking any value after optimization that violates $g_i(x) \leq 0$ as it will have an infinite cost ensuring the unconstrained optimization problem indirectly satisfy $g_i(x) \leq 0$.
 
\begin{figure} [!htbp]  	 	
 	\includegraphics[width=1\textwidth]{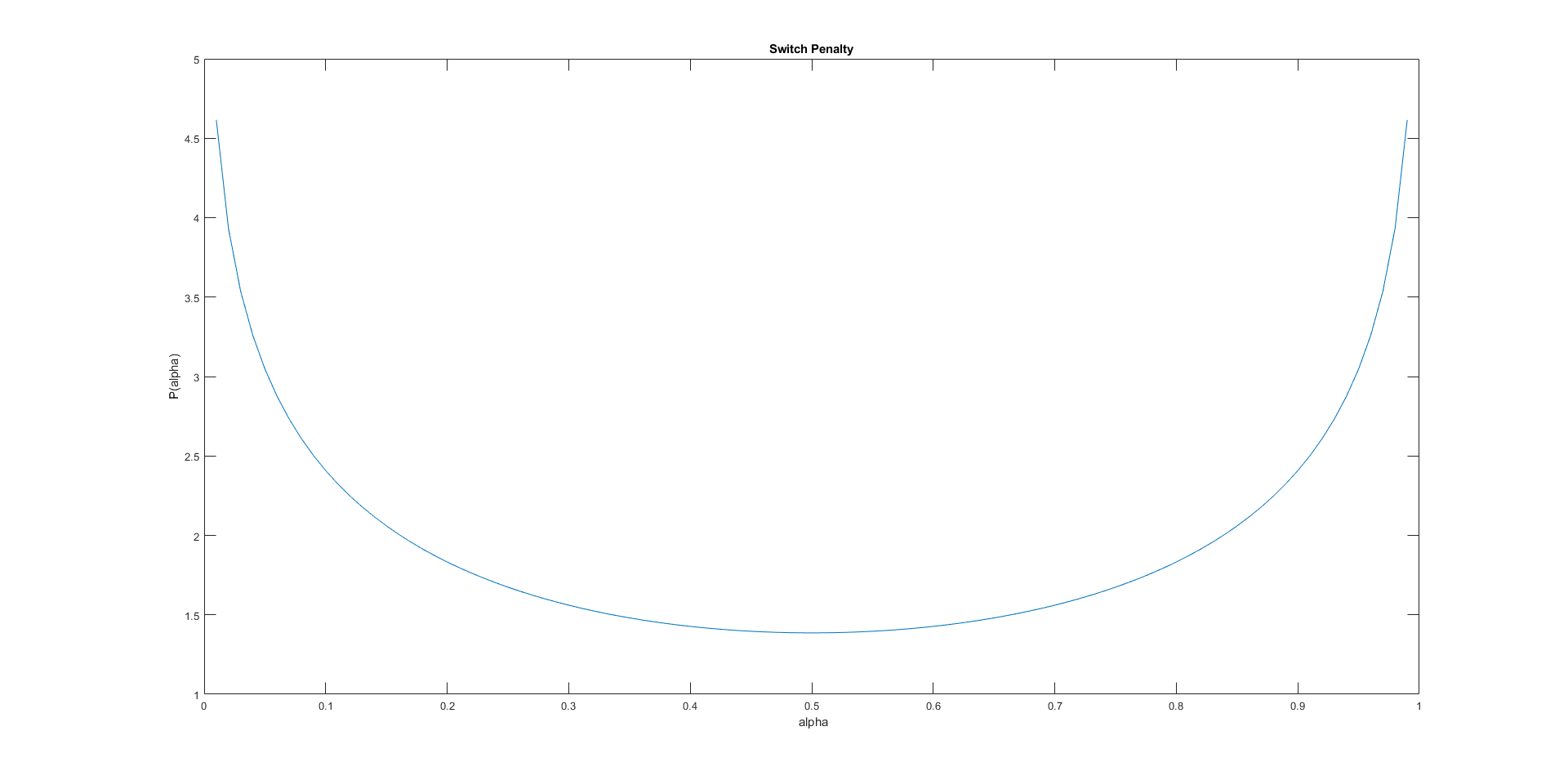}
 	\caption{The switch penalty}
 	\label{Image: Switch_Penalty} 	
 \end{figure}

 Now coming back to our optimization problem, $\alpha_s$ is restricted to take values in the set $[0\;\;1]$. This could be equivalently posed as the following 2 inequalities:
 \begin{align}
 	\alpha_s \geq 0 \\
 	\alpha_s \leq 1
 \end{align}
 Equivalently it can also be stated in standard inequality form as:
 \begin{align}
 	-\alpha_s \leq  0 \\
 	\alpha_s-1 \leq 0
 \end{align} 
 If we had to construct a $\log$ barrier to ensure the above $2$ inequalities, it would be of the following form: 
 \begin{equation}
 	\text{ Barrier :}-\log(\alpha_s)-\log(1-\alpha_s)
 \end{equation} 
 By appealing to the property that $\log(ab)=\log(a)+\log(b)$, we obtain a simplified form:
 \begin{equation}
 	\text{ Barrier :}-\log\big((\alpha_s).(1-\alpha_s)\big)
 \end{equation} 
 Hence the term, $P_{\tau}( {\alpha_s})= -\tau \log\big(\alpha_s(1-\alpha_s)\big)$ can be understood to encourage $\alpha_s$ to stay in the interval $[0\;\;1]$ after optimization.Here $\tau$, is a scaling parameter of the barrier function  which influences how strongly this constraint is imposed. The term $P_{\tau}( {\alpha_s})$ also serves a second role which was reported in literature \cite{corosa}  for a variant of this barrier function for the  n-dimensional weight case. The second purpose  is to discourage switching of local weight parameters which is valid here as well.   The  ability to discourage switching is  evident from the following  \Cref{Image: Switch_Penalty} of the term  $P_{\tau}( {\alpha_s})= -\log\big(\alpha_s(1-\alpha_s)\big)$ as a function of $\alpha_s$.  
 
 It may be seen that as $\alpha$ moves closer to $1$ , the cost function records a nonlinear increase. Similarly as as $\alpha$ moves closer to $0$, we see a similar nonlinear increase in the cost associated with $P_{\tau}( {\alpha})$. In other words $\alpha$ is encouraged to be close to $0.5$ ensuring spurious switching does not happen. The non-linearity of this increase means, that farther you are away, higher the cost. This ensures that extreme values are more strongly discouraged ensuring intermediate weights are also permitted as solution.
 
 \subsection{Action of $\alpha_s$ on two terms in our regularization}
 Let us take a closer look at partial cost function with  only the image regularization term :
 \begin{align*}   
 	{\alpha_s}\bigg(\sum_{i=1}^{n_F}\sum_{\*k}\|A_s\big((\mathbf{T}_s*\*f_i) \big)\|_{2}\bigg) + 	(1- {\alpha_s })\bigg( \|(\mathbf{T_t}*\*f)  \|_{1,F}\bigg)  
 \end{align*}
 A simplified expression emphasizing the pixel wise action  of weights $(\alpha_s)$ and $(1-\alpha_s)$  will take the following form:
 \begin{align*}   
 	\sum_{i=1}^{n_F}\sum_{\mathbf{k}} \bigg(	{\alpha_s }\|A_s\big((\mathbf{T}_s*\*f_i)(\*k) \big)\|_{F}  + 	 \ (1- {\alpha_s })\|(\mathbf{T_t}*\*f_i)(\*k)  \|_{F} \bigg)  
 \end{align*}
 We can observe that in the PSTAIC model, the same weight  $(\alpha_s)$ act on all spatial points of the first term in regularization. Similarly the same  weight  $(1-\alpha_s)$ act  on all spatial points  in the second term of the regularization. In other words, the weights are chosen the same for all spatial points. The needed spatial motion aware behavior is achieved through the the infimal decomposition of the signal that happens locally as a result of the infimal convolution based definition of our STAIC formulation. This  is a basic difference in the action of PSTAIC model as against a sum of norms regularization model with spatially varying weights. This is because   sum of norms approach to image  regularization have the same effect at all pixel positions (not spatially aware) which necessitates need for spatially variation in weights.

 \section{Algorithm for PSTAIC Model Optimization} \label{Chapter_4:PSTAIC_ALGORITHM}
 
 The optimization of the PSTAIC cost $H(\*f,\alpha_s)$ is performed through an alternating minimization scheme based on the { COROSA}  \cite{corosa} algorithm . The optimization in  weight vector  $ {\alpha_s}$ is performed by assuming a fixed image  $\*f$ first. This is followed by updation of   $\*f$ by assuming  $\alpha_s$ is fixed. The resultant  optimization  scheme is  summarized  below:

 \begin{algorithm}[H]
 	Initialization : $\*f^{(0)}$\\
  	\text{For }{$\ell=0 \;\;$ to $\;\;$  \text{N\_total}  }\\
 		$\text{   }\quad {\alpha_s}^{(\ell+1)} = \argmin{0 \leq \alpha_s \leq 1}{H(\*f^{(\ell)}, {\alpha_s})}$\\
 		$\text{   }\quad \mathbf{f}^{(\ell+1)} = \argmin{\*f}{H(\*f, {\alpha_s}^{(\ell+1)})}$
 	
 	\caption{PSTAIC Iterative Scheme }
 \end{algorithm}
 We can observe two important smaller optimization problems in variable $\alpha_s$ and $\*f$ alone as stated below:
 \begin{enumerate}
 	\item The weight sub-problem :
 	$ \argmin{0 \leq \alpha_s \leq 1}{H(\*f^{(\ell)}, {\alpha_s})}$ 
 	\item The image  sub-problem :
 	$ \argmin{\*f}{H(\*f, {\alpha_s}^{(\ell+1)})}$ 
 \end{enumerate}

 The solution of optimization problems  in variables $\alpha_s$ and  $\*f$        is discussed   in  \Cref{Section : Weight_Subproblem} and     \Cref{Section :PSTAIC_Image_Subproblem} respectively.

 			\subsection{The Weights sub-problem to solve for $\alpha_s$} 
 			\label{Section : Weight_Subproblem}
 			
 			Now, we consider the first sub-problem involving the   weight  $ {\alpha_s}$ while assuming that $\*f$ takes a fixed value obtained in the previous step. The optimization problem involving only ${\alpha_s}$ may be described as follows: 
 			\begin{equation*}\small
 				{\alpha_s}^{(\ell+1)}  =\argmin{ {0 
 						\leq \alpha_s \leq 1}}{	H(\*f^{(\ell)}, {\alpha_s} )} 
 			\end{equation*} 
 			where $\*f^{(\ell)}$ is the image estimate obtained in the previous step of iteration. The expanded form of the optimization problem is given by:  
 			\begin{align*}
 				{\alpha_s}^{(\ell+1)}  = \underset{0 
 					\leq \alpha_s \leq 1}{ \operatorname{argmin\; }} &\frac{1}{2}\sum_{i=1}^{n_F}\|(\*h*\mathbf{f}_i^{(\ell)} ) -m_i \|_{1,F}^2 +\\
 				&\lambda \big(\sqrt{2} {\alpha_s } \sum_{i=1}^{n_F}\sum_{\*k}\|A_s\big(\mathbf{T_s}*\*f_i^{(\ell))}(\*k) \big)\|_{F}  +   (1- {\alpha_s })\|(\mathbf{T_t}*\*f^{(\ell)})  \|_{1,F}\big)  +	\mathcal{B_C}(\*f^{(\ell)})+P_{\tau}( {\alpha_s})
 			\end{align*}

 			It may be observed that $\mathbf{f}^{(\ell)}$  is a constant    in this equation.  This allows us to ignore constant terms not dependent  on $\alpha_s$ in the above expression  and retain only functions involving the variable ${\alpha_s}$. The updated optimization problem in ${\alpha_s}$ 
 			\begin{equation}\small
 				{\alpha_s}^{(\ell+1)} 
 				=\argmin{ {0\leq{\alpha_s}\leq 1}}{\lambda \big( {\alpha_s } \sqrt{2}\sum_{i=1}^{n_F}\sum_{\*k}\|A_s\big(\mathbf{T_s}*\*f_i^{(\ell)})(\*k) \big)\|_{F}  +   (1- {\alpha_s })\|(\mathbf{T_t}*\*f^{(\ell)})  \|_{1,F}\big) +P_{\tau}( {\alpha_s}) }
 			\end{equation} 
 			Now we expand the definition of the term $P_{\tau}( {\alpha_s})$ to obtain a simplified version of the weight sub-problem as:
 			\begin{equation}\label{Equation: Weight_Optimziation_Problem}
 				{\alpha_s}^{(\ell+1)} 
 				=\argmin{ {0\leq{\alpha_s}\leq 1}}{\lambda \big( \sqrt{2}{\alpha_s } \sum_{i=1}^{n_F}\sum_{\*k}\|A_s\big(\mathbf{T_s}*\*f_i^{(\ell)}) (\*k)\big)\|_{F}  +   (1- {\alpha_s })\|(\mathbf{T_t}*\*f^{(\ell)})  \|_{1,F}\big)  -\tau \log(\alpha_s (1-\alpha_s )) }
 			\end{equation}

 			The solution of the above optimization problem can be  derived by appealing to 	( Proposition 1 . \cite{corosa}) by making necessary changes to the coefficients of function of $\alpha_s$. 
 			
 			\begin{lemma}
 				[Based on Proposition 1 \cite{corosa}]	The solution of the optimization problem in \Cref{Equation: Weight_Optimziation_Problem} is given by   $\alpha_s^+=\frac{1}{2}(\sqrt{\zeta^2+1}-\zeta)$ and $\alpha_s^-=\frac{1}{2}(-\sqrt{\zeta^2+1}-\zeta)$   where $\zeta = \frac{2\tau}{|C_1-C_2|}$. Here    $C_1=\sqrt{2}\lambda \sum_{i=1}^{n_F}\sum_{\*k}  \|A_s(\mathbf{T_s}*\*f_i^{(\ell)})(\*k) \|_{  2}$ and $C_2=\lambda \|(\mathbf{T_t}*\*f^{(\ell)}) \|_{1, 2}$ .
 			\end{lemma}
 			By accepting only the positive solution of $\alpha_s$, we can conclude that the optimal $\alpha_s$ is obtained as follows:

 			\begin{equation}
 				{\alpha_s}^{(\ell+1)} =
 				\begin{cases}
 					\frac{1}{2}\bigg( 1-sign(\mu)\big(\sqrt{\frac{4\tau^2}{\mu^2}+1}-\frac{2\tau}{|\mu|}\big)\bigg) \text{ when }  \mu\neq 0 \\
 					\frac{1}{2}  \text{ when  } \mu= 0 
 				\end{cases}
 			\end{equation}  
 			
 			where $\mu =C_1-C_2$. Here, we have used the definition of $\zeta = \frac{2\tau}{|C_1-C_2|}$ to get an expression only in terms of $\mu$.

 			\subsubsection{Behavior of the weights sub-problem in relation to its parameters}
 			An analysis of the solution of the  weight sub-problem gives us insights into how the weights are updated after each alternating step involving variables $\*f$ and $ {\alpha_s}$. Recalling the  formula for the weight update step we obtained that:
 			
 			\begin{equation}
 				{\alpha_s}=\frac{1}{2}\bigg( 1-sign(\mu )\big(\sqrt{\frac{4\tau^2}{\mu^2 }+1}-\frac{2\tau}{|\mu |}\big)\bigg) \text{when }  \mu \neq 0
 			\end{equation}
 			
 			Also $\alpha_s=\frac{1}{2} \text{ when } \mu=0$. It may be observed that  
 			$\sqrt{\frac{4\tau^2}{\mu^2 }+1} > \frac{2\tau}{|\mu |}$ because $\sqrt{\frac{4\tau^2}{\mu^2 }}= \frac{2\tau}{|\mu |}$. From above it may be concluded that $\frac{1}{2}\bigg( 1-sign(\mu )\big(\sqrt{\frac{4\tau^2}{\mu^2 }+1}-\frac{2\tau}{|\mu |}\big)\bigg)$ will take a  value bigger than 0.5 if $\mu<0$ and smaller than 0.5 if $\mu>0$.

 			\subsubsection{Properties of weight sub-problem cost}
 			The first important observation  that can be made about the weight sub-problem in \Cref{Equation: Weight_Optimziation_Problem} is the differentiability of the cost in variable $\alpha_s$. A closer look at the weight sub-problem  enables us to identify the following properties that helps us understand the behavior of the weights obtained as a solution of the weights sub-problem.
 			
 			\begin{enumerate}
 				\item  The weight sub-problem cost is strictly convex.
 				\item A closed form solution exists for the weight sub-problem.
 			\end{enumerate}
 			
 			The first  property has to do with the convexity of the weight sub-problem cost . It can be shown to be a strictly convex function in relation to the variable ${\alpha_s}$ as proved  in the following theorem. 
 			
 			\begin{theorem}
 				The cost function 	$H(\*f, {\alpha_s} )$ evaluated at      $\*f^{(\ell)}$ is strictly convex function in variable $ {\alpha_s}$
 			\end{theorem}
 			
 			\begin{proof}
 				The cost function $H(\*f, {\alpha_s} )$ evaluated at the point $\*f^{(\ell)}$ may be stated as follows : 	
 				\begin{align*}
 					H(\*f^{(\ell)}, {\alpha_s} )&=	\frac{1}{2}\sum_{i=1}^{n_F}\|(\*h*\mathbf{f}_i^{(\ell)})(\*k)-m_i(\*k)\|_{1,2}^2\\ 
 					&+\bigg(  {\alpha_s}\lambda \sum_{i=1}^{n_F}\sum_{\*k}  \|A_s(\mathbf{T_s}*\*f_i^{(\ell)})(\*k) \|_{ 2} +   (1- {\alpha_s} )\lambda \|(\mathbf{T_t}*\*f^{(\ell)}) \|_{ 2}\bigg) +	\mathcal{B_C}({\*f}^{(\ell)})+P_{\tau}( {\alpha_s} )
 				\end{align*} 	
 				Ignoring the constant terms independent of $ {\alpha_s}$ and using the definition $P( {\alpha_s} , \tau)=-  \tau \log( {\alpha_s}(1- {\alpha_s})$, the optimization cost may be restated as follows: 
 				\begin{align*}
 					\bar{H}( {\alpha_s})=   {\alpha_s}\lambda \sum_{i=1}^{n_F}\sum_{\*k}  \|A_s(\mathbf{T_s}*\*f_i^{(\ell)})(\*k) \|_{  2}+  (1- {\sqrt{2}\alpha_s})\lambda \ \|(\mathbf{T}_a*\*f^{(\ell)})(\*r)\|_{1, 2}  -  \tau \log( {\alpha_s}(1- {\alpha_s})) 
 				\end{align*}
 				By defining  	  $C_1=\lambda \sum_{i=1}^{n_F}\sum_{\*k}  \|A_s(\mathbf{T_s}*\*f_i^{(\ell)})(\*k) \|_{  2}$ and $C_2=\lambda \|(\mathbf{T_t}*\*f^{(\ell)}) \|_{1, 2}$ , we obtain a simplified form of the weight cost (represented by $	\bar{H}( {\alpha_s})$) :  	
 				\begin{equation}
 					\bar{H}( {\alpha_s})= {\alpha_s} C_1  +   (1- {\alpha_s})C_2  -  \tau \log( {\alpha_s}(1- {\alpha_s})) 
 				\end{equation}  	
 				The derivative of $\bar{H}( {\alpha_s})$ with respect to variable $\alpha_s $ is given by : 	
 				\begin{equation}
 					\frac{ \mathrm{d}	\bar{H} }{\mathrm{d} {\alpha_s}  }= 	(C_1-C_2)-\frac{\tau}{\alpha_s } + \frac{\tau}{1-\alpha_s }
 				\end{equation} 	
 				The second derivative of $\bar{H}( {\alpha_s})$ with respect to variable $\alpha_s$ is given by : 	
 				\begin{equation}
 					\frac{ \mathrm{d}^2	\bar{H} }{\mathrm{d} {\alpha_s^2}  }=	 \frac{\tau}{\alpha_s ^2} + \frac{\tau}{(1-\alpha_s )^2}
 				\end{equation} 	
 				Since $\tau>0$, 	$\frac{\partial^2 	\bar{H}( {\alpha_s})}{\partial  {\alpha_s}^2  } \geq 0$ for $0< {\alpha_s}<1$. 	
 				Recall that a strictly convex function will have a positive second derivative by definition of strict convexity. 
 				Hence the cost function 	$H(\*f, {\alpha_s} )$ evaluated at      $\*f^{(\ell)}$ is strictly convex function in variable $\alpha_s$ for $ 0<{\alpha_s}<1$.
 			\end{proof}
 			
 			The above theorem helps us conclude the following:
 			\begin{enumerate}
 				\item[I)] The weight sub-problem cost is a  strictly convex function.
 				\item[II)] The weight sub-problem in \Cref{Equation: Weight_Optimziation_Problem} have a global unique solution in $(0\;\;1)$.
 			\end{enumerate}
 			This permits us to confidently look for the global optima in $(0\;\;1)$ which  will give us the intermediate estimate of the weight parameter.

 					\subsubsection{Behavior of Estimated weights as a function of $\mu=C_1-C_2$}
 					
 					From previous description of formula for $\alpha_s$, we can observe that $\alpha_s(\mu)$ ($\alpha_s$ as a function of $\mu$) has a two part    definition due to its behavior at $\mu=0$. At $\mu=0$, the function is specifically assigned a value $\alpha_s(\mu)=\frac{1}{2}$.   This raises a question whether the function is continuous at $\mu=\frac{1}{2}$.  
 					Continuity at $\mu=\frac{1}{2}$ is a good property to have as it will mean that weights will not change dramatically as we move from $\mu=0+\epsilon$  to $\mu=0-\epsilon$ where $\epsilon$ is a small positive value near $0$. It turns out that the function $\alpha_s(\mu)$ is indeed a continuous function at $\mu=0$. This is proved in the following theorem.

 					\begin{theorem}
 						The function   $\alpha_s(\mu)$ is  continuous function at $\mu =0$.
 					\end{theorem}
 					\begin{proof}
 						
 						Recall that a  function $f(x): \mathbb{R} \rightarrow \mathbb{R}$ is said to be continuous at a point c if 
 						\begin{enumerate}
 							\item  $\lim_{x \to c}$  exists and ,
 							\item 	$\lim_{x \to c} f(x) =f(c)$
 						\end{enumerate}
 						To show that the $\displaystyle \lim_{\mu \to 0}\alpha_s(\mu)$ exists, we needs to prove that the left hand limit is equal to right hand limit. 	We first consider the left hand limit of the given function
 						\begin{equation}
 							\lim_{\mu\to 0^-} \alpha_s(\mu)   =\lim_{\mu\to 0^-} \frac{1}{2}\bigg( 1-sign(\mu )\big(\sqrt{\frac{4\tau^2}{\mu^2 }+1}-\frac{2\tau}{|\mu |}\big)\bigg)
 						\end{equation}
 						Since $\mu\to 0^-$, by using substitution $\mu=0-h$ where $h>0$ it can be equivalently stated as:
 						\begin{equation}
 							\lim_{h\to 0} \frac{1}{2}\bigg( 1-sign(0-h)\big(\sqrt{\frac{4\tau^2}{h^2 }+1}-\frac{2\tau}{|-h|}\big)\bigg)=\lim_{h\to 0} \frac{1}{2}\bigg( 1+\big(\sqrt{\frac{4\tau^2}{h^2 }+1}-\frac{2\tau}{h}\big)\bigg)
 						\end{equation}
 						
 						By appealing to the 	L'Hôpital's rule we get the following limit as $0$
 						\begin{equation}
 							\lim_{h\to 0}  \big(\sqrt{\frac{4\tau^2}{h^2 }+1}-\frac{2\tau}{h}\big) = \lim_{h\to 0} \frac{2h}{2\sqrt{4\tau^2+h^2}}=0
 						\end{equation}
 						Hence 	$\lim_{\mu\to 0^-} \alpha_s(\mu)=\frac{1}{2}$. 	
 						
 						We next consider the right hand limit of the given function
 						\begin{equation}
 							\lim_{\mu\to 0^+} \alpha_s(\mu)   =\lim_{\mu\to 0^+} \frac{1}{2}\bigg( 1-sign(\mu )\big(\sqrt{\frac{4\tau^2}{\mu^2 }+1}-\frac{2\tau}{|\mu |}\big)\bigg)
 						\end{equation}
 						Since $\mu\to 0^+$,by using substitution $\mu=0+h$ where $h>0$ it can be it can be equivalently stated as:
 						\begin{equation}
 							\lim_{h\to 0} \frac{1}{2}\bigg( 1-sign(0+h)\big(\sqrt{\frac{4\tau^2}{h^2 }+1}-\frac{2\tau}{|h|}\big)\bigg)=\lim_{h\to 0} \frac{1}{2}\bigg( 1-\big(\sqrt{\frac{4\tau^2}{h^2 }+1}-\frac{2\tau}{h}\big)\bigg)=\frac{1}{2}
 						\end{equation}  
 						Hence 	$\lim_{\mu\to 0^+} \alpha_s(\mu)=\frac{1}{2}$.
 						
 						In summary $\displaystyle \lim_{\mu\to 0^-} \alpha_s(\mu)=\lim_{\mu\to 0^+} \alpha_s(\mu)$  and $\displaystyle \lim_{\mu\to 0} \alpha_s(\mu)=\alpha_s(0)=\frac{1}{2}$.
 						Finally, can conclude that $\alpha_s(\mu)$ is continuous at $\mu=\frac{1}{2}$. 	 
 					\end{proof}

 					This is important as it confirms that the weights are not radically different close to the place where the difference in regularization switches signs.

 					\subsubsection{Effect of $\tau $  on weight sub-problem}
 					
 					The parameter $\tau$ is a user defined parameter that controls the strength of the switching penalty in the weight sub-problem.  The weight $ {\alpha_s}$ is understood to be switched if it changes from a value smaller than $0.5$ to a value greater than $0.5$ as it goes from one iteration to the next. A change from a value greater than $0.5$ to a value smaller than $0.5$ is also interpreted as switched. A high values of $\tau$ discourages a value of  $ {\alpha_s}$ that is close to either $0$ or $1$. In other words it encourages a slower update of $ {\alpha_s}$ from one iteration to next.  A smaller value of $\tau$  may lead to  extreme values of $ {\alpha_s}$ which may be unfavorable for a stable convergence towards an optimal weight. This value of $\tau$ is commonly chosen for all images in the experiment.

 					\subsection{ Image Sub-problem to Solve for  $\*f$} 
 					\label{Section :PSTAIC_Image_Subproblem}
 					
 					In this subsection, we focus on the sub-problem involving the variable $\*f$ which we refer to as the image sub-problem  as $\*f$ represents the  spatio-temporal image being estimated.      
 					The sub-problem to be solved in variable $\*f$ may be compactly written as follows:      
 					\begin{equation}\label{Equation:PSTAIC:Weight Subproblem}
 						\mathbf{f}^{(\ell+1)}  = \argmin{\*f}{H(\*f, {\alpha_s} ^{(\ell+1)})}
 					\end{equation} 
 					where  
 					\begin{align*}\small
 						H(\*f, {\alpha_s} ^{(\ell+1)} ) = &\frac{1}{2}\sum_{i=1}^{n_F}\|(\*h*\mathbf{f}_i) -m_i \|_{1,F}^2+ \\
 						&\lambda   {\sqrt{2}\alpha_s } ^{(\ell+1)} \sum_{i=1}^{n_F}\sum_{\*k}\|A_s\big((\mathbf{T_s}*\*f_i)(\*k) \big)\|_{F}  +  \lambda (1- {\alpha_s } ^{(\ell+1)})\|(\mathbf{T_t}*\*f)  \|_{1,F}  +\mathcal{B_C}(\*f)+P_{\tau}( {\alpha_s} ^{(\ell+1)})
 					\end{align*} 
 					Here $ {\alpha_s} ^{(\ell+1)}$ is the value from the previous step of the algorithm. It can be treated as a constant in the context of above sub-problem.
 					This enables us to ignore the term $P_{\tau}( {\alpha_s} ^{(\ell+1)})$ as it do not involve the variable of optimization $\*f$ and can be treated as a constant in optimization cost. Hence the updated image sub-problem cost is:
 					\begin{align*}\small
 						H(\*f, {\alpha_s} ^{(\ell+1)} ) = &\frac{1}{2}\sum_{i=1}^{n_F}\|(\*h*\mathbf{f}_i) -m_i \|_{1,F}^2+ \\
 						&\lambda  {\sqrt{2}\alpha_s } ^{(\ell+1)} \sum_{i=1}^{n_F}\sum_{\*k}\|A_s\big((\mathbf{T_s}*\*f_i)(\*k) \big)\|_{F}  +   \lambda(1- {\alpha_s } ^{(\ell+1)})\|(\mathbf{T_t}*\*f)  \|_{1,F}  +\mathcal{B_C}(\*f) 
 					\end{align*} 
 					Before presenting an algorithm for minimizing this cost, we demonstrate the convexity of sub-problem cost function in $\*f$ alone in the following lemma.

 						\begin{theorem} \label{Lemma: Convexity of Hf} 
 							The cost function 	$H(\*f, {\alpha_s} )$ evaluated at $ {\alpha_s} ^{(\ell+1)}$  is a   convex function in variable $\*f$.
 						\end{theorem}
 						
 						\begin{proof}
 							
 							The cost function $H(\*f, {\alpha_s} )$ evaluated at the point $ {\alpha_s} ^{(\ell+1)}$ may be stated as follows :	
 							\begin{align*}\small
 								H(\*f, {\alpha_s} ^{(\ell+1)} ) = &\frac{1}{2}\sum_{i=1}^{n_F}\|(\*h*\mathbf{f}_i) -m_i \|_{1,F}^2+ \\
 								&\lambda \sqrt{2} {\alpha_s } ^{(\ell+1)} \sum_{i=1}^{n_F}\sum_{\*k}\|A_s\big((\mathbf{T_s}*\*f_i)(\*k)\big)\|_{F}  +   \lambda(1- {\alpha_s } ^{(\ell+1)})\|(\mathbf{T_t}*\*f)  \|_{1,F}  +\mathcal{B_C}(\*f) 
 							\end{align*} 
 							We will prove convexity of the full cost by analyzing the individual components in the sum.
 							
 							The data fitting term $\frac{1}{2}\sum_{i=1}^{n_F}\|(\*h*\mathbf{f}_i )(\*k)-m_i(\*k)\|_{F}^2$ is convex as it is a sum of convex functions(norms) .   $	\mathcal{B}({\*f} )$ is indicator function of a convex set. Hence  $	\mathcal{B}({\*f} )$ is a convex function. The term $P( {\alpha_s} ^{(\ell+1)} , \tau)$ is a constant term with respect to $\*f$.	The term $ \lambda \sqrt{2}{\alpha_s} ^{(\ell+1)} \sum_{\*r}   \|A_s(\mathbf{T_s}*\*f_i )(\*r)\|_{ 2} $ is a sum of convex functions. Since sum of convex functions are also convex, this term must also must be convex. Here  	the weights of summation $\alpha_s ^{(\ell+1)} >0$. Similarly $\lambda (1- {\alpha_s} ^{(\ell+1)} )\sum_{\*r}\|(\mathbf{T_t}*\*f )(\*r)\|_{ 2}$ is also a convex function. since sum of convex functions are also convex, this term must also be convex. Here  	the weights of summation $(1-\alpha_s  ^{(\ell+1)})>0$ since we restrict the weights to be in $[0 \;\; 1]$. Since $ 	H(\*f , {\alpha_s} ^{(\ell+1)} )$ is a sum of convex functions it  must also be   convex in variable $\*f$ when ${\alpha_s} ^{(\ell+1)}$ takes a value from the previous iteration .	
 						\end{proof}
 						
 						The convexity of   	$H(\*f, {\alpha_s} ^{(\ell+1)} )$ is exploited to design an algorithm to minimize it by using the ADMM iterative scheme.     As a first step the convex optimization  problem must be  reformulated to get an equivalent  linearly constrained convex problem so that its fits well into the ADMM framework for algorithm design.
 						
 						\subsubsection{ Image Sub-problem : Reformulation into a Linearly Constrained Problem }
 						We introduce new variables   $\mathbf{w}_s,\mathbf{w}_t,\*w_b$ and $w_m$ and rewrite the optimization problem in \Cref{Equation:PSTAIC:Weight Subproblem} as follows: 
 						\begin{align*}
 							&\operatornamewithlimits{min}\limits_{\mathbf{f},\*w_s,\*w_t,\*w_b,w_m}{\;\;\frac{1}{2}\sum_{i=1}^{n_F}\|w_{m_i} -m_i\|_{1,F}^2} + \lambda \sqrt{2}  {\alpha_s } ^{(\ell+1)} \sum_{i=1}^{n_F}\sum_{\*k}\|A_s ( \mathbf{w}_s(\*k)  )\|_{F}  +   \lambda (1- {\alpha_s } ^{(\ell+1)})\|\mathbf{w}_t \|_{1,F}  +\mathcal{B_C}(\mathbf{w}_b)\\
 							&\text{ subject to }  
 							\;\;\;\;\;\;\;\;\;\;\mathbf{h}*\mathbf{f}_i=w_{m_i},\;\;\mathbf{T}_s*\mathbf{f}_i=\mathbf{w}_s,\;\;\mathbf{T_t}*\mathbf{f}=\mathbf{w}_t,\;\;  \*f=\mathbf{w}_b 
 						\end{align*}

 						It may be noted that we have converted the problem to a constrained form with only linear equality constraints.  To allow a simpler algorithm statement, we introduce a combined operator $\mathbf{T}(\*r)$ and a combined vector $\*w$ defined as follows :
 						
 						\begin{align}
 							&\*T(\*r) = \begin{bmatrix}
 								\*h(\*r) \\
 								\*T_s(\*r) \\
 								\*T_t(\*r) \\
 								\*e(\*r)
 							\end{bmatrix},
 							&\*w=\begin{bmatrix}
 								w_m \\
 								\*w_s \\
 								\*w_t \\
 								\*w_b
 							\end{bmatrix}
 						\end{align}
 						where $\mathbf{e}(\*r)=[\delta(\*r) \;\; \delta(\*r)]$ where $\delta(\*r)$ denotes Kronecker delta. Under this definition, $\mathbf{e} * \*f = \*f $.
 						The linearly constrained problem can now  be stated in a compact form as: 
 						\begin{align}
 							&	(\mathbf{f}^*,\mathbf{w}^*) = \argmin{\*f,\*w}{R(\*w,\alpha_s^{(\ell+1)})}  \\
 							& \text{ subject to } \quad \quad\mathbf{T}*\*f = \*w   \nonumber
 						\end{align}   
 						where $\displaystyle R(\*w,\alpha_s^{(\ell+1)})=\frac{1}{2}\sum_{i=1}^{n_F}\|w_{m_i} -m_i\|_{1,F}^2 + \lambda \sqrt{2}   {\alpha_s } ^{(\ell+1)} \sum_{i=1}^{n_F}\sum_{\*k}\|A_s\big( \mathbf{w}_s(\*k) \big)\|_{F} +   \lambda (1- {\alpha_s } ^{(\ell+1)})\|\mathbf{w}_t \|_{1,F}  +\mathcal{B_C}(\mathbf{w}_b)) $. The next step in ADMM framework is to construct the Augmented Lagrangian  \cite{boyd2011distributed}  $\mathcal{L}(\*f,\*w,{\pmb \beta},\alpha_s ^{(\ell+1)})$  of the above linearly constrained cost.    
 						\begin{equation} \small
 							\mathcal{L}(\*f,\*w,{\pmb\beta},\alpha_s ^{(\ell+1)}) = R(\*w,\alpha_s ^{(\ell+1)}) + \langle {\pmb \beta}, \mathbf{T}*\mathbf{f}-\mathbf{w}\rangle +\frac{\rho}{2}\|\mathbf{T}*\mathbf{f}-\mathbf{w}\|_2^2
 						\end{equation}
 						where ${\pmb \beta} $ is the Lagrangian multiplier and $\rho$ is a user supplied ADMM parameter. Here the dimensions of $\pmb \beta$ is same as that of $\*w$. Finally, ADMM algorithm involves collection of alternative minimization of sub-problem with respect to variables $\*f$ and $\*w$ followed by an update step involving variable $\pmb \beta$. Assume that $\*f^k,\*w^k ,{\pmb \beta}^k$ are the current estimates , the ADMM algorithm involves the following steps
 						
 						\begin{align}   		  		
 							\*w^{k+1}&=\operatornamewithlimits{argmin}\limits_{\*f}{{L}(\*f^{(k)},\*w,{\pmb \beta}^{(k)},\alpha_s ^{(\ell+1)})} \label{Equation : PSTAIC_w-sub-problem}\\
 							\*f^{k+1}&=\operatornamewithlimits{argmin}\limits_{\*f}{L(\*f,\*w^{(k+1)},{\pmb \beta}^k,\alpha_s ^{(\ell+1)})} \label{Equation : PSTAIC_f-sub-problem}\\
 							\text{ and } \;\;{\pmb \beta}^{(k+1)}& = {\pmb \beta}^{(k)} + \rho\big(\mathbf{T}*\mathbf{f}^{(k+1)}-\mathbf{w}^{(k+1)}\big) \label{Equation : PSTAIC_lambda-sub-problem}
 						\end{align}
 						
 						The first two equation involves solving two optimization problems over variables $\mathbf{f}$ and $\mathbf{w}$ respectively which is discussed next. All the three components of the iterative scheme follows construction similar to ADMM sub-problems in STAIC formulation.

 						\section{Solving the Sub-problems of ADMM Algorithm} \label{Section : ADMM subproblems}
 						
 						We will now discuss how the  sub-problems in   \Cref{Equation : PSTAIC_w-sub-problem} and \Cref{Equation : PSTAIC_f-sub-problem} are solved to obtain the intermediate variables  
 						$\*w^{(k+1)} $ and $\*f^{(k+1)}$  that appear in the PSTAIC iterative scheme.
 						It may be noted that a lot of the solutions in this subsection is similar to the previous paper, we include it for completeness.
 						
 						\subsection{The w problem}
 						The sub-problem cost in   \Cref{Equation : PSTAIC_w-sub-problem} can be equivalently simplified to the following form    
 						\begin{align}
 							&	L_{w,k}(\*w,\alpha_s^{\ell+1},1-\alpha_s^{\ell+1})=  R(\*w,\alpha_s^{\ell+1},1-\alpha_s^{\ell+1}) + \frac{\rho}{2}\|\*w-\bar{\*x}^{(k)}\|_2^2 \\
 							&\quad \text{ where }  \quad 
 							\bar{\mathbf{x}}^{(k)} = \mathbf{T}*{\*f^{(k)}} + \frac{1}{\rho}{\pmb \beta^{(k)}}
 						\end{align}
 						For cleaner presentation of sub-problems, we introduce the notation $\*x=\bar{\*x}^{(k)}$ and $\hat{\*w}=\*w^{(k+1)}$.  Since $\*w$ is made up of sub vectors $w_m,\*w_b,\*w_t,\*w_s$, we separate the above problem into sub-problems involving constituent variables.
 						
 						\begin{align}
 							\label{eq:wmprob2}
 							\mbox{$w_m$-prob.:} \; &  \hat{w}_m = 
 							\underset{w_m}{\operatorname{argmin}} 
 							\underbrace{\frac{\rho}{2} \|x_m-w_m\|_{2,2}^2 + 
 								G({w}_m,   {m})}_{\bar{L}_{m}(w_m,x_m)}  \\
 							\label{eq:wbprob2}
 							\mbox{$\*w_b$-prob.:} \;\;\;&  \hat{\*w}_b = 
 							\underset{\*w_b}{\operatorname{argmin}} \;\; 
 							\underbrace{\frac{\rho}{2} \|\*x_b-\*w_b\|_{2,2}^2 + 
 								{\mathcal{B_C}}(\*w_b)}_{\bar{L}_{b}(\*w_b,\*x_b)}  \\
 							\label{eq:wfprob2}
 							\mbox{${\mathbf w}_t$-prob.:} \;\;\; &   \hat{\mathbf w}_t = 
 							\underset{{\mathbf w}_t}{\operatorname{argmin}} \;\; 
 							\underbrace{\frac{\rho}{2} \|{\mathbf x}_t-{\mathbf w}_t\|_{2,2}^2 + 
 								(1-\alpha_s^{\ell+1}) \| \mathbf{w}_t\|_{1,2}}_{\bar{L}_{t}({\mathbf w}_t, {\mathbf x}_t, 1-\alpha_s^{\ell+1})} \\
 							\label{eq:wsprob2}
 							\mbox{${ \*w}_s$-prob.:} \;\;\; &  \hat{  \*w}_s =  
 							\underset{{\mathbf w}_s}{\operatorname{argmin}} \;\; \sum_{i=1}^{n_F}
 							\underbrace{\frac{\rho}{2} \|{\mathbf x}_{s_i}-{\mathbf w}_{s_i}\|_{2,2}^2 + 
 								\sqrt{2}\alpha_s^{\ell+1}  \sum_{\mathbf{k}}\|{A}_s (\mathbf{w}_{s_i}(\mathbf{k}))\|_{2}}_{\bar{L}_{s_i}({\mathbf w}_{s_i}, 
 								{\mathbf x}_{s_i}, \alpha_s^{\ell+1})} 
 						\end{align}
 						\subsubsection{Decomposing problems pixel-wise}
 						The cost $\bar{L}_{m}(w_m,x_m)$ is separable across pixels as shown below:
 						
 						\begin{align}
 							\bar{L}_{m}(w_m,x_m)  = & 
 							\frac{\rho}{2} \|x_m-w_m\|_{2,2}^2 + 
 							G({w}_m,m)  \\
 							= &   \frac{\rho}{2} \|x_m-w_m\|_{2,2}^2 +    \frac{1}{2}\| w_m -m\|_{2}^2\\
 							=&\sum_{\*r} \underbrace{ \frac{\rho}{2} (x_m(\*r)-w_m(\*r))^2 +    \frac{1}{2}(w_m(\*r) -m(\*r))^2}_{{L}_{m}(w_m(\r),x_m(\r))}
 						\end{align}

 						Hence  the pixel wise cost ${L}_{m}(w_m(\r),x_m(\r))$ is given by:
 						\begin{equation*}
 							{L}_{m}(w_m(\r),x_m(\r))  =     \frac{\rho}{2} (x_m(\r)-w_m(\r))^2   
 							+   \frac{1}{2}(w_m(\r)-m(\r))^2
 						\end{equation*}

 						The cost function $\bar{L}_{b}(w_b,x_b)$ is separable across 3D pixel index $\*r$ because
 						$\mathcal{B_C}(\mathbf{w}_b)= \sum_{\*r} \bar{ \mathcal{B_C}}(\mathbf{w}_b(\*r)) $ where  
 						\begin{equation}
 							\bar{ \mathcal{B_C}}(\mathbf{w}_b(\*r)) =
 							\begin{cases}
 								\text{0} &\quad\text{if } \mathbf{w}_b(\*r) \geq 0\\
 								\infty &\quad\text{otherwise} 
 							\end{cases}
 						\end{equation}
 						The cost function reformulated as a sum over pixels can be stated as:
 						\begin{equation}
 							\bar{L}_{b}(w_b,x_b)   = \sum_\*r 
 							\underbrace{\frac{\rho}{2} (\mathbf{x}_b(\*r)-\mathbf{w}_b(\*r))^2 + 
 								\bar{ \mathcal{B_C}}(\mathbf{w}_b(\*r))}_{
 								{L}_{b}(\mathbf{w}_b(\*r),\mathbf{x}_b(\*r)) }
 						\end{equation}

 						Now $\bar{L}_{s_i}({\mathbf w}_{s_i}, {\mathbf x}_{s_i}, \alpha_s^{\ell+1})$ can be expanded across pixels as follows:
 						\begin{equation}
 							\bar{L}_{s_i}({\mathbf w}_{s_i}, {\mathbf x}_{s_i}, \alpha_s^{\ell+1})  = \sum_{\mathbf{k}}
 							\underbrace{\frac{\rho}{2}\|{\mathbf x}_{s_i}(\mathbf{k})-{\mathbf w}_{s_i}(\mathbf{k})\|_{2}^2 + 
 								 	\lambda \sqrt{2}\alpha_s^{\ell+1} \|A_s{  \*w}_{s_i}(\mathbf{k})\|_{{2}}}_{
 								{L}_{s}({\mathbf w}_{s_i}(\mathbf{k}), {\mathbf x}_{s_i}(\mathbf{k}), \alpha_s^{\ell+1})},
 						\end{equation}
 						
 						Finally, $\bar{L}_{t}({\mathbf w}_t, {\mathbf x}_t, 1-\alpha_s^{(\ell+1)})$ can also be expanded across pixels courtesy the use of mixed vector matrix norm
 						
 						\begin{equation} \label{Equation: ws_subproblem_pixel}
 							\bar{L}_{t}({\mathbf w}_t, {\mathbf x}_t, 1-\alpha_s^{(\ell+1)})    = \sum_\*r
 							\underbrace{
 								\frac{\rho}{2} \|{\mathbf x}_t(\*r)-{\mathbf w}_t(\*r)\|_{2}^2 + 
 								\lambda (1-\alpha_s^{\ell+1}) \| {\mathbf w}_t(\*r)\|_{2}}_{
 								{L}_{t}({\mathbf w}_t(\*r), {\mathbf x}_t(\*r), 1-\alpha_s^{(\ell+1)})  }
 						\end{equation}

 						We have shown so far that all  the sub-problems are separable across pixels.
 						Hence the solution to the
 						minimization problems of equations \eqref{eq:wmprob2},   \eqref{eq:wbprob2}, 
 						\eqref{eq:wsprob2},
 						and \eqref{eq:wfprob2}, can be expressed as following:
 						\begin{align}
 							\label{eq:wmrobpix}
 							\hat{w}_m(\r) = & 
 							\underset{z\in \mathbb{R}}{\operatorname{argmin}} \;\;
 							{L}_{m}(z,x_m(\r)), \\
 							\label{eq:wbrobpix}
 							\hat{\mathbf{w}}_b(\r) = & 
 							\underset{z\in \mathbb{R}}{\operatorname{argmin}} \;\;
 							{L}_{b}(z,\mathbf{x}_b(\r)), \\
 							\label{eq:wsprobpix}
 							\hat{\mathbf w}_{s_i}(\mathbf{k}) = &
 							\underset{{\mathbf z}\in \mathbb{R}^{10}}{\operatorname{argmin}} \;\;
 							{L}_{s}({\mathbf z},{\mathbf x}_{s_i}(\mathbf{k}), \alpha_s^{\ell+1})\text{ , and}  \\
 							\label{eq:wfprobpix}
 							\hat{\mathbf w}_t(\r) = &
 							\underset{{\mathbf z}\in \mathbb{R}^9}{\operatorname{argmin}} \;\;
 							{L}_{t}({\mathbf z},{\mathbf x}_t(\r), 1-\alpha_s^{\ell+1}).
 						\end{align}
 						\subsubsection{{Solution to  the pixel-wise sub-problems}}
 						\label{sec:subp}
 						
 						The solution to the $ {w}_m$ sub-problem is  obtained  by exploiting the fact that the cost  ${L}_{m}(z,x_m(\r))$ is a differentiable function. The minima is obtained by  finding the stationary point of the cost function  and the resultant optimal point $\hat{w}_m(\r)$ is:
 						\begin{equation}
 							\hat{w}_m(\r)  = \frac{\rho x_m(\r)  + m(\r)}{\rho + 1}
 						\end{equation}
 						
 						The solution to the
 						$w_b$-problem  is also  simple, and it is the clipping of the pixels by bound
 						that defines the set $\mathcal{C}$  \cite{parikh2014proximal}.
 						The optimal point  $\hat{w}_b(\r)$  as given below:
 						\begin{equation}
 							\hat{\mathbf{w}}_b(\r) =  \mathbb{P}_{ \mathcal{C}}(\mathbf{x}_b(\r)),
 						\end{equation}
 						where ${\mathbb{ P}}_{\mathcal{C}}(\cdot)$  denotes the operation of clipping the pixel values within the   bounds in definition of $\mathcal{C}$.

 						The $\mathbf{w}_t$ sub-problem could be understood as evaluating the well known proximal operator \cite{parikh2014proximal} of $\ell_2$ norm $\frac{\lambda(1-\alpha_s^{\ell+1})}{\rho}\|\cdot \|_2$ at the point ${\mathbf x}_t(\*r)$.  Hence the solution to the  $\mathbf{w}_t$ sub-problem is given by 
 						
 						\begin{align}
 							\hat{\mathbf w}_t(\r)&= 	\underset{{\mathbf z}\in \mathbb{R}^9}{\operatorname{argmin}} \;\;	\frac{\rho}{2} \|{\mathbf x}_t(\*r)-\*z\|_{2}^2 + 
 							\lambda(1-\alpha_s^{\ell+1}) \| {\mathbf z} \|_{2}\\
 							&=\begin{cases}
 								\big(1-\frac{\lambda(1-\alpha_s^{\ell+1})}{\rho \|{\mathbf x}_t(\*r)\|_2}\big){\mathbf x}_t(\*r) & \|{\mathbf x}_t(\*r)\|_2\geq \frac{\lambda(1-\alpha_s^{\ell+1})}{\rho}\\
 								\mathbf{0} &otherwise
 							\end{cases}
 						\end{align}
 						
 						The $\mathbf{w}_s$ sub-problem is more complicated as its involves composition of  a linear operator with a norm function. The solution of this sub-problem is given by the following lemma :

 						\begin{lemma} \cite{skariah2024staic} \label{Lemma : As proximal} Let $\mathbf{y} \in \mathbb{R}^{10}$. Let $\mathbf{y}_1\in \mathbb{R}^{5}$ be its sub-vector with first five entries and $\mathbf{y}_2\in \mathbb{R}^{5}$ be its sub-vector with last five entries. The solution of optimization problem  $\;\;\argmin{\mathbf{z} \in \mathbb{R}^{10}}{\frac{\rho}{2}\| {{  \mathbf{y}}}-\mathbf{z}\|_{2}^2 + 
 								\sqrt{2} \lambda	\alpha_s^{\ell+1} \|A_s \mathbf{z}\|_{{2}}}\;\;$  where   $A_s$ is defined in  \eqref{Definition : Matrix A_s} is given by
 							\begin{equation} 	\mathbf{z}^* = P\left[\begin{array}{c} \gamma {{\mathbf{y}}}_{1} \\ {{\mathbf{y}}}_{2}\end{array} \right] \end{equation}
 							Here    $P$  defined in \eqref{Defintion : Matrix  P of eigenvectors} is the eigenvector matrix of $A_s$  and  $\gamma =  \max(0,1 -\frac{\lambda \sqrt{2} \alpha_s^{\ell+1}}{\rho \|{{\mathbf{y}}}_{1}\|_2})$.
 						\end{lemma}
 						\begin{equation} \label{Defintion : Matrix  P of eigenvectors}
 							P= \left[ \begin{array}{cccccccccc}
 								\frac{1}{\sqrt{2}} & 0 & 0 & 0 & 0 &-\frac{1}{\sqrt{2}} &0 &0 &0 & 0 \\
 								0 &\frac{1}{\sqrt{2}} & 0 & 0 & 0 & 0 &-\frac{1}{\sqrt{2}} &0 &0 &0   \\
 								0 & 0 &\frac{1}{\sqrt{2}} & 0 & 0 & 0 & 0 &-\frac{1}{\sqrt{2}} &0 &0    \\
 								0&0 & 0 &\frac{1}{\sqrt{2}} & 0 & 0 & 0 & 0 &-\frac{1}{\sqrt{2}} &0    \\
 								0&	  0&0 & 0 &\frac{1}{\sqrt{2}} & 0 & 0 & 0 & 0 &-\frac{1}{\sqrt{2}}     \\
 								-\frac{1}{\sqrt{2}} & 0 & 0 & 0 & 0 &\frac{1}{\sqrt{2}} &0 &0 &0 & 0 \\
 								0 &-\frac{1}{\sqrt{2}} & 0 & 0 & 0 & 0 &\frac{1}{\sqrt{2}} &0 &0 &0   \\
 								0 & 0 &-\frac{1}{\sqrt{2}} & 0 & 0 & 0 & 0 &\frac{1}{\sqrt{2}} &0 &0    \\
 								0&0 & 0 &-\frac{1}{\sqrt{2}} & 0 & 0 & 0 & 0 &\frac{1}{\sqrt{2}} &0    \\
 								0&	  0&0 & 0 &-\frac{1}{\sqrt{2}} & 0 & 0 & 0 & 0 &\frac{1}{\sqrt{2}}     \\
 							\end{array}  \right]
 							\end{equation}
 						By appealing to this lemma, we can conclude that solution of $\mathbf{w}_s$ sub-problem is as follows:
 						\begin{equation}
 							\hat{\mathbf{w}}_{s_i}(\*k)= P\left[\begin{array}{c} \gamma \mathbf{x}_{s_{i1}}(\*k)  \\ \mathbf{x}_{s_{i2}}(\*k) \end{array} \right]
 						\end{equation}
 						
 						where $\mathbf{x}_{s_{i1}}(\*k) \in \mathbb{R}^5$ is the sub-vector of $\mathbf{x}_{s_{i}}(\*k) $  with first five entries and  $\mathbf{x}_{s_{i2}}(\*k) \in \mathbb{R}^5$ is the sub-vector  of $\mathbf{x}_{s_{i}}(\*k) $ with last five entries.

 						\subsection{The f sub-problem }
 						\begin{equation}	
 							\*f^{(k+1)}=\operatornamewithlimits{argmin}\limits_{\*f}{L(\*f,\*w^{(k+1)},{\pmb \beta}^{(k)},\alpha_s^{\ell+1},1-\alpha_s^{\ell+1})}  
 						\end{equation}

 						Thesub-problem in variable 	  $\*f $  given in  \eqref{Equation : PSTAIC_f-sub-problem}   has a simpler form once you ignore all the terms not depending on $\mathbf{f}$ in the optimization problem. The simpler from of $\mathbf{f}$ sub-problem may be stated as follows:
 						
 						\begin{eqnarray}
 							\*f^{(k+1)} =&\operatornamewithlimits{argmin}\limits_{\*f}\frac{1}{2}\|\mathbf{T}*\mathbf{f}-\mathbf{y}^{(k)}\|_2^2 \\\text{ where } 
 							\nonumber	& \mathbf{y}^{(k)}=\mathbf{w}^{(k+1)}-\frac{1}{\rho} \pmb{\beta}^{(k)}
 						\end{eqnarray}

 						For notational convenience, we let $\mathbf{y} = \mathbf{y}^{(k)}$ and $\hat{\mathbf{f}}=\mathbf{f}^{(k+1)}$.
 						Recall that $\mathbf{f(\*r)}=[g(\*r) \;\;v(\*r)]^\top$. From the definition of  $\mathbf{T}(\*r)$ it can be observed that the cost is separable across the components $g$ and $v$ of $\mathbf{f}(\*r)$. Assume that $\mathbf{y}(\*r) =
 						[ {y}_m(\*r)\;  {y}_{s,1}(\*r) \;  {y}_{s,2}(\*r)\;   {y}_{s,3}(\*r)\;   {y}_{s,4}(\*r)\;   {y}_{s,5}(\*r)\;  {y}_{s,6}(\*r)  \;  {y}_{s,7}(\*r)\;  {y}_{s,8}(\*r) \;  {y}_{s,9}(\*r)\;  {y}_{s,10}(\*r)\;    {y}_{t,1}(\*r)\; \\ {y}_{ t,2}(\*r)\;  {y}_{t,3}(\*r)\;   {y}_{t,4}(\*r)\;   {y}_{t,5}(\*r)\;  {y}_{t,6}(\*r)\;  {y}_{t,7}(\*r)\;  {y}_{t,8}(\*r)\;  {y}_{t,9}(\*r)  \;   {y}_{b,1}(\*r)\;   {y}_{b,2}(\*r)]^\top$. This simplification is achieved by observing  the structure  $\mathbf{y}^{(k)}$ it inherits from $\mathbf{w}^{(k+1)}$ and $\pmb{\beta}^{(k)}$. Now the cost separated along $g$ and $v$ is given by:
 						\begin{align*}
 							\nonumber   \bar{L}_1(g)  = &  \frac{1}{2} \big(\|h*g - y_{m}\|_{2}^2 
 							+  \|d_{xx}*g - y_{s,1}\|_{2}^2+  \|d_{yy}*g - y_{s,2}\|_{2}^2 +\\&
 							\|d_{xy}*g - y_{s,3}\|_{2}^2 +	\|d_{xy}*g - y_{s,4}\|_{2}^2+  \|g - y_{s,5}\|_{2}^2+\|g - y_{b,1}\|_{2}^2 \big)
 						\end{align*} 
 						\begin{align*}
 							\nonumber   \bar{L}_2(v)  =  & \frac{1}{2} \big(\|d_{xx}*v - y_{s,6}\|_{2}^2 
 							+   \|d_{yy}*v - y_{s,7}\|_{2}^2+ \|d_{xy}*v - y_{s,8}\|_{2}^2 +\|d_{xy}*v - y_{s,9}\|_{2}^2 + \\& 
 							\|v - y_{s,10}\|_{2}^2 +\|d_{xx}*v - y_{t,1}\|_{2}^2 
 							+   \|d_{yy}*v - y_{t,2}\|_{2}^2+ \|d_{xy}*v - y_{t,3}\|_{2}^2 + \\& \|d_{xy}*v - y_{t,4}\|_{2}^2
 							\|d_{yt}*v - y_{t,5}\|_{2}^2 +\|d_{yt}*v - y_{t,6}\|_{2}^2+  \|d_{xt}*v - y_{t,7}\|_{2}^2+\\&\|d_{xt}*v - y_{t,8}\|_{2}^2+ \|d_{tt}*v - y_{t,9}\|_{2}^2 +\|v - y_{b,2}\|_{2}^2 \big) 
 						\end{align*}

 						For notational convenience, we let $\mathbf{y} = \mathbf{y}^{(k)}$ and $\hat{\mathbf{f}}=\mathbf{f}^{(k+1)}$.
 						Recall that $\mathbf{f(\*r)}=[g(\*r) \;\;v(\*r)]^\top$

 						The function $ \bar{L}_1(g)$ and $ \bar{L}_2(v)$ are quadratic in nature in the variables $g$ and $v$ respectively. The minima of both these functions can be obtained by solving the equations$\nabla_{g}\bar{L}_1(g)=\mathbf{0}$ and $\nabla_{v}\bar{L}_2(v)=\mathbf{0}$ respectively. This requires evaluation of the gradient expressions which are given below:
 						
 						\begin{align}
 							\nabla_{g}\bar{L}_1(g) = & \tilde{h} *h*g +\tilde{d}_{xx}*d_{xx}*g+\tilde{d}_{yy}*d_{yy}*g+2\tilde{d}_{xy}*d_{xy}*g+g\\\nonumber &-\tilde{h}*y_m	- \tilde{d}_{xx}*y_{s,1}   -\tilde{d}_{yy}*y_{s,2}-\tilde{d}_{xy}*y_{s,3}-\tilde{d}_{xy}*y_{s,4}-y_{s,5}-y_{b,1}
 						\end{align}
 						\begin{align}
 							\nabla_{v}\bar{L}_2(v) = & \tilde{d}_{xx}*d_{xx}*v+ \tilde{d}_{yy}*d_{yy}*v+2\tilde{d}_{xy}*d_{xy}*v+2v
 							+\tilde{d}_{xt}*d_{xt}*v\\ \nonumber&+\tilde{d}_{yt}*d_{yt}*v+\tilde{d}_{tt}*d_{tt}*v 
 							- \tilde{d}_{xx}*y_{s,5} -\tilde{d}_{yy}*y_{s,6}-\tilde{d}_{xy}*y_{s,7}-\tilde{h}*y_m-y_{s,8} \\ \nonumber&
 							- \tilde{d}_{xt}*y_{t,1} -\tilde{d}_{yt}*y_{t,2}-\tilde{d}_{tt}*y_{t,3}  
 							- \tilde{d}_{xt}*y_{t,4} -\tilde{d}_{yt}*y_{t,5}-\tilde{d}_{tt}*y_{t,6}
 						\end{align}

 						\section{ Experiments}   \label{Chapter_4:EXPERIMENTS}
 						We designed an algorithm that performs joint estimation of the restored spatio-temporal TIRF image along with the weights that controls the relative importance of  spatial against temporal smoothing at each pixel point. We considered a set of  of 5 TIRF signals named Image 5135,  Image 5142, Image 5147, Image  5157 and Image 5158.  
 						We then simulate measured images by blurring these models
 						with PSF corresponding to low NA systems and by adding mixed Poisson-Gaussian noise as shown below :
 						\begin{equation*}
 							m = \mathcal{P}(\gamma_p(h * g))  + \eta 
 						\end{equation*} 
 						Here  $\gamma_p$ is a  parameter to control the strength of Poisson noise and our dataset was generated by setting $\gamma_p=1$.
 						We consider PSF corresponding to  five NA values namely 0.8, 0.9, 1.0, 1.1,  and 1.2. 
 						
 						\begin{figure*}[!htbp]
 							\centering
 							\includegraphics[width=.7\textwidth]{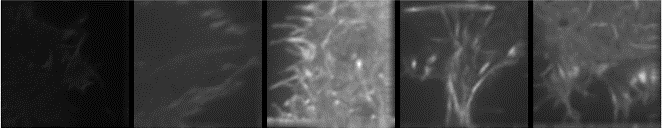}
 							\caption{ The five original   Images }
 							\label{Figure:Fiveimages}
 						\end{figure*}
 					  We compare the performance of the PSTAIC algorithm against weight estimation equipped ICTV  where we strengthen the ICTV \cite{holler2014infimal} using the same algorithmic approach to parameter selection. We call this weight estimation enabled ICTV algorithm as  PICTV. The PICTV sub-problem is similar to the sub-problems of PSTAIC except that the regularization terms involved are inherited from ICTV.

 						\subsection{Experiments 1}
 						\begin{table}[!htbp]
 					
 							\begin{tabular}{|cl|ll|ll|}
 								\hline
 								\multicolumn{2}{|c|}{\textbf{}}                                                                                                          & \multicolumn{2}{c|}{\textbf{PSTAIC}}                                                                                          & \multicolumn{2}{c|}{\textbf{PICTV}}                                                                                           \\ \hline
 								\rowcolor[HTML]{FFFFFF} 
 								\multicolumn{1}{|c|}{\cellcolor[HTML]{FFFFFF}\textbf{Image}}                  & \multicolumn{1}{c|}{\cellcolor[HTML]{FFFFFF}\textbf{NA}} & \multicolumn{1}{c|}{\cellcolor[HTML]{FFFFFF}\textbf{ssim}} & \multicolumn{1}{c|}{\cellcolor[HTML]{FFFFFF}\textbf{SNR(in dB)}} & \multicolumn{1}{c|}{\cellcolor[HTML]{FFFFFF}\textbf{ssim}} & \multicolumn{1}{c|}{\cellcolor[HTML]{FFFFFF}\textbf{SNR(in dB)}} \\ \hline
 								\multicolumn{1}{|c|}{\cellcolor[HTML]{FFFFFF}}                                & \cellcolor[HTML]{FFFFFF}0.8                              & \multicolumn{1}{l|}{0.823}                                 & 9.67                                                             & \multicolumn{1}{l|}{0.786}                                 & 8.74                                                             \\ \cline{2-6} 
 								\multicolumn{1}{|c|}{\cellcolor[HTML]{FFFFFF}}                                & \cellcolor[HTML]{FFFFFF}0.9                              & \multicolumn{1}{l|}{0.842}                                 & 10.83                                                            & \multicolumn{1}{l|}{0.801}                                 & 9.80                                                             \\ \cline{2-6} 
 								\multicolumn{1}{|c|}{\cellcolor[HTML]{FFFFFF}}                                & \cellcolor[HTML]{FFFFFF}1                                & \multicolumn{1}{l|}{0.855}                                 & 11.85                                                            & \multicolumn{1}{l|}{0.810}                                 & 10.72                                                            \\ \cline{2-6} 
 								\multicolumn{1}{|c|}{\cellcolor[HTML]{FFFFFF}}                                & \cellcolor[HTML]{FFFFFF}1.1                              & \multicolumn{1}{l|}{0.865}                                 & 12.75                                                            & \multicolumn{1}{l|}{0.815}                                 & 11.52                                                            \\ \cline{2-6} 
 								\multicolumn{1}{|c|}{\multirow{-5}{*}{\cellcolor[HTML]{FFFFFF}\textbf{5135}}} & \cellcolor[HTML]{FFFFFF}1.2                              & \multicolumn{1}{l|}{0.872}                                 & 13.57                                                            & \multicolumn{1}{l|}{0.817}                                 & 12.24                                                            \\ \hline
 								\multicolumn{1}{|c|}{\cellcolor[HTML]{FFFFFF}}                                & \cellcolor[HTML]{FFFFFF}0.8                              & \multicolumn{1}{l|}{0.903}                                 & 10.75                                                            & \multicolumn{1}{l|}{0.877}                                 & 10.13                                                            \\ \cline{2-6} 
 								\multicolumn{1}{|c|}{\cellcolor[HTML]{FFFFFF}}                                & \cellcolor[HTML]{FFFFFF}0.9                              & \multicolumn{1}{l|}{0.916}                                 & 11.96                                                            & \multicolumn{1}{l|}{0.888}                                 & 11.25                                                            \\ \cline{2-6} 
 								\multicolumn{1}{|c|}{\cellcolor[HTML]{FFFFFF}}                                & \cellcolor[HTML]{FFFFFF}1                                & \multicolumn{1}{l|}{0.924}                                 & 13.05                                                            & \multicolumn{1}{l|}{0.894}                                 & 12.25                                                            \\ \cline{2-6} 
 								\multicolumn{1}{|c|}{\cellcolor[HTML]{FFFFFF}}                                & \cellcolor[HTML]{FFFFFF}1.1                              & \multicolumn{1}{l|}{0.931}                                 & 14.04                                                            & \multicolumn{1}{l|}{0.898}                                 & 13.14                                                            \\ \cline{2-6} 
 								\multicolumn{1}{|c|}{\multirow{-5}{*}{\cellcolor[HTML]{FFFFFF}\textbf{5142}}} & \cellcolor[HTML]{FFFFFF}1.2                              & \multicolumn{1}{l|}{0.935}                                 & 14.96                                                            & \multicolumn{1}{l|}{0.898}                                 & 13.96                                                            \\ \hline
 								\multicolumn{1}{|c|}{\cellcolor[HTML]{FFFFFF}}                                & \cellcolor[HTML]{FFFFFF}0.8                              & \multicolumn{1}{l|}{0.907}                                 & 12.25                                                            & \multicolumn{1}{l|}{0.913}                                 & 11.93                                                            \\ \cline{2-6} 
 								\multicolumn{1}{|c|}{\cellcolor[HTML]{FFFFFF}}                                & \cellcolor[HTML]{FFFFFF}0.9                              & \multicolumn{1}{l|}{0.925}                                 & 13.41                                                            & \multicolumn{1}{l|}{0.927}                                 & 13.03                                                            \\ \cline{2-6} 
 								\multicolumn{1}{|c|}{\cellcolor[HTML]{FFFFFF}}                                & \cellcolor[HTML]{FFFFFF}1                                & \multicolumn{1}{l|}{0.938}                                 & 14.45                                                            & \multicolumn{1}{l|}{0.936}                                 & 14.02                                                            \\ \cline{2-6} 
 								\multicolumn{1}{|c|}{\cellcolor[HTML]{FFFFFF}}                                & \cellcolor[HTML]{FFFFFF}1.1                              & \multicolumn{1}{l|}{0.947}                                 & 15.38                                                            & \multicolumn{1}{l|}{0.942}                                 & 14.91                                                            \\ \cline{2-6} 
 								\multicolumn{1}{|c|}{\multirow{-5}{*}{\cellcolor[HTML]{FFFFFF}\textbf{5147}}} & \cellcolor[HTML]{FFFFFF}1.2                              & \multicolumn{1}{l|}{0.953}                                 & 16.24                                                            & \multicolumn{1}{l|}{0.946}                                 & 15.71                                                            \\ \hline
 								\multicolumn{1}{|c|}{\cellcolor[HTML]{FFFFFF}}                                & \cellcolor[HTML]{FFFFFF}0.8                              & \multicolumn{1}{l|}{0.859}                                 & 7.86                                                             & \multicolumn{1}{l|}{0.837}                                 & 7.19                                                             \\ \cline{2-6} 
 								\multicolumn{1}{|c|}{\cellcolor[HTML]{FFFFFF}}                                & \cellcolor[HTML]{FFFFFF}0.9                              & \multicolumn{1}{l|}{0.883}                                 & 9.27                                                             & \multicolumn{1}{l|}{0.858}                                 & 8.50                                                             \\ \cline{2-6} 
 								\multicolumn{1}{|c|}{\cellcolor[HTML]{FFFFFF}}                                & \cellcolor[HTML]{FFFFFF}1                                & \multicolumn{1}{l|}{0.900}                                 & 10.54                                                            & \multicolumn{1}{l|}{0.872}                                 & 9.67                                                             \\ \cline{2-6} 
 								\multicolumn{1}{|c|}{\cellcolor[HTML]{FFFFFF}}                                & \cellcolor[HTML]{FFFFFF}1.1                              & \multicolumn{1}{l|}{0.912}                                 & 11.67                                                            & \multicolumn{1}{l|}{0.882}                                 & 10.70                                                            \\ \cline{2-6} 
 								\multicolumn{1}{|c|}{\multirow{-5}{*}{\cellcolor[HTML]{FFFFFF}\textbf{5157}}} & \cellcolor[HTML]{FFFFFF}1.2                              & \multicolumn{1}{l|}{0.920}                                 & 12.71                                                            & \multicolumn{1}{l|}{0.888}                                 & 11.63                                                            \\ \hline
 								\multicolumn{1}{|c|}{\cellcolor[HTML]{FFFFFF}}                                & \cellcolor[HTML]{FFFFFF}0.8                              & \multicolumn{1}{l|}{0.939}                                 & 14.09                                                            & \multicolumn{1}{l|}{0.929}                                 & 13.12                                                            \\ \cline{2-6} 
 								\multicolumn{1}{|c|}{\cellcolor[HTML]{FFFFFF}}                                & \cellcolor[HTML]{FFFFFF}0.9                              & \multicolumn{1}{l|}{0.950}                                 & 15.43                                                            & \multicolumn{1}{l|}{0.938}                                 & 14.31                                                            \\ \cline{2-6} 
 								\multicolumn{1}{|c|}{\cellcolor[HTML]{FFFFFF}}                                & \cellcolor[HTML]{FFFFFF}1                                & \multicolumn{1}{l|}{0.958}                                 & 16.65                                                            & \multicolumn{1}{l|}{0.943}                                 & 15.36                                                            \\ \cline{2-6} 
 								\multicolumn{1}{|c|}{\cellcolor[HTML]{FFFFFF}}                                & \cellcolor[HTML]{FFFFFF}1.1                              & \multicolumn{1}{l|}{0.963}                                 & 17.75                                                            & \multicolumn{1}{l|}{0.946}                                 & 16.29                                                            \\ \cline{2-6} 
 								\multicolumn{1}{|c|}{\multirow{-5}{*}{\cellcolor[HTML]{FFFFFF}\textbf{5158}}} & \cellcolor[HTML]{FFFFFF}1.2                              & \multicolumn{1}{l|}{0.967}                                 & 18.76                                                            & \multicolumn{1}{l|}{0.948}                                 & 17.13                                                            \\ \hline
 							\end{tabular}
 								
 							\caption{Parameter Selection in PSTAIC and PICTV}
 								\label{Table: PICTV and PSTAIC}
 						\end{table}
 				 
 						In the PICTV algorithm used in comparison, we equip the ICTV scheme with a  weight variable that is also part of the optimization problem. Formally , the PICTV optimization problem takes the following form:
 						
 						\begin{equation} 
 							ICTV(g) = 	\operatornamewithlimits{min} \limits_{v}\left[\alpha_1\|\nabla(g-v)  \|_{1,\kappa_1}  + 
 							\alpha_2\|\nabla(v) \|_{1,1/\kappa_2}\right],
 						\end{equation}

 						In the above formulation, we restrict $\alpha_2=1-\alpha_1$ and add it as part of the optimization problem. This leads us to a formulation similar to the the PSTAIC formulation involving the weights variable and the image variable. This new formulation of ICTV we refer to as PICTV. We apply this algorithm on the given dataset of 25 different noisy images by tuning for the best SNR over the regularization parameter $\lambda$. The results are provided in the attached \Cref{Table: PICTV and PSTAIC}.

 						In the experiments for PSTAIC we choose the regularization parameter $\lambda$ that provides the best SNR value. One design choice to be made is the value of the parameter $\tau$. We chose to treat $\tau$ as a spatial parameter where we use $e^{-D_{tt}f_0^2}$ where $f_0$ is the previous estimate. It ensures that we discourage fast $\pmb{\alpha}$ parameter change if there is presence of motion in that region.  The results of the experiments are presented in \Cref{Table: PICTV and PSTAIC}. It can be observed that PSTAIC is better than PICTV in majority cases  of simulated images in terms of  both SSIM and snr measures.   
 						\section{Conclusion}
 						
 						We proposed a model for joint estimation of parameters and   image for   restoration of TIRF spatio-temporal images based on the STAIC regularization scheme. We also designed an alternating minimization scheme to minimize the proposed restoration cost. We demonstrated the superior reconstruction quality of the proposed scheme as against weight estimation enabled ICTV scheme. 
 						 
\bibliographystyle{unsrtnat}
\bibliography{citation}  
\end{document}